# Accelerating the prediction of inorganic surfaces with machine learning interatomic potentials

Kyle Noordhoek[1], Christopher J. Bartel[1]*

[1]Department of Chemical Engineering and Materials Science, University of Minnesota, Minneapolis, MN, USA 55455

*correspondence to cbartel@umn.edu

## Abstract

The surface properties of solid-state materials often dictate their functionality, especially for applications where nanoscale effects become important. The relevant surface(s) and their properties are determined, in large part, by the material's synthesis or operating conditions. These conditions dictate thermodynamic driving forces and kinetic rates responsible for yielding the observed surface structure and morphology. Computational surface science methods have long been applied to connect thermochemical conditions to surface phase stability, particularly in the heterogeneous catalysis and thin film growth communities. This review provides a brief introduction to first-principles approaches to compute surface phase diagrams before introducing emerging data-driven approaches. The remainder of the review focuses on the application of machine learning, predominantly in the form of learned interatomic potentials, to study complex surfaces. As machine learning algorithms and large datasets on which to train them become more commonplace in materials science, computational methods are poised to become even more predictive and powerful for modeling the complexities of inorganic surfaces at the nanoscale.

## Introduction

Surface science and nanoscale synthesis are key driving factors in many current technological applications including catalysis[1] and microelectronics.[2] For catalysis applications, surface reactivity is dictated by the structure of exposed surfaces on nanoparticles or thin films. Understanding the phase stability of relevant surfaces is therefore paramount for catalyst design. In thin-film devices, interfacial interactions between substrates and vapor-deposited materials dictate phase stability and, again, the observed properties are highly dependent upon the surface or interfacial structure. Hence, accurately capturing which surfaces are likely to be observed under relevant conditions plays an important role in the design of nanostructured solid-state materials.

Before we can discuss the broader methods for understanding surface stability, it is necessary to introduce the terminology commonly used to distinguish plausible surfaces. An inorganic surface is modeled as a slab – an infinite 2D sheet of material formed by slicing a bulk (3D) crystal using a particular 2D plane. The cleavage of the conventional unit cell through a designated Miller plane produces a single facet. Surface facets are nominally referred to using Miller index notation to indicate the plane used to perform the slice with respect to the conventional unit cell. A facet (Miller index) alone does not define a slab as the position in the 3D crystal where the cut is made can lead to different "terminations" of the slab (i.e., different atomic species at the surface). It is typical for multiple possible terminations per facet to be generated when computing surface properties. For a more detailed description of how surface slabs with varying terminations can be systematically generated as a starting point for first-principles calculations, see the thorough explanation given by Sun and Ceder.[3] After the generation of a facet with a particular termination, the "dangling bonds" formed by slicing the bulk material can induce a rearrangement of atomic positions at/near the surface in a process known as reconstruction. Reconstructions are often denoted using Wood's notation,[4] which describes modifications of the surface unit cell compared with the bulk (e.g., the well-known 7×7 reconstruction of Si).[5,6] Understanding the surface structure is critical for countless applications, and the relative energies of these various reconstructions, facets, and terminations determine which surface structures are likely to appear for a material at a given set of conditions (**Figure 1**). This review focuses primarily on recent efforts to use machine learning to address the challenging problem of calculating the thermodynamics of solid-state, inorganic surfaces using first-principles methods.

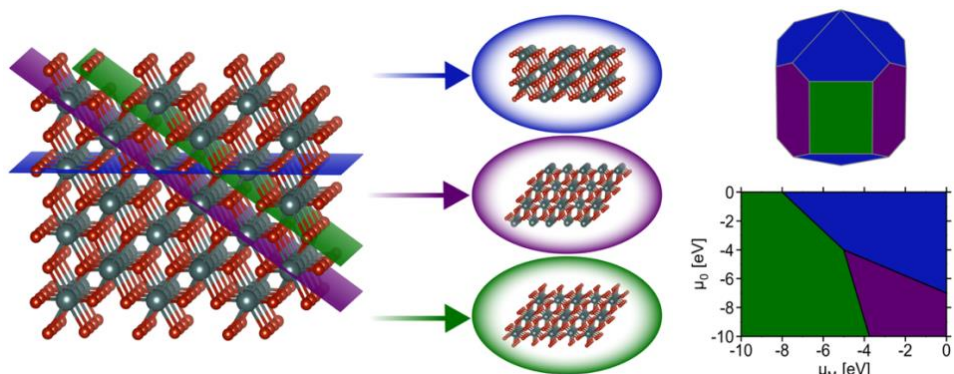

**Figure 1.** Illustrating how a 3D crystal can be cleaved by 2D planes to yield various slabs, which are used as the starting point for surface science calculations. As an illustrative example, we consider a (001) facet (blue) and two terminations of the (011) facet (purple, green). Once the surface energies are known, they can be used as inputs to the Wulff construction to yield an equilibrium nanoparticle geometry (top right) or thermodynamic models to understand how the stability of each facet depends on the chemical potentials of the involved elements (bottom right for a monometallic metal oxide).

## Computational thermodynamics of surfaces

The surface (internal) energy, γ, of a slab in vacuum can be computed as the difference between the total internal energy of the slab, $E_{slab}$, and the total internal energy of the bulk, $E_{bulk}$, given the same number of atoms, $N$, as the slab. For the case where the upper and lower slab surfaces are identical, the surface free energy is calculated as:

$$\gamma = \frac{1}{2A} * (E_{slab} - NE_{bulk})  \quad\quad [1]$$

where $A$ is the area of the surface and $2A$ arises from the two identical surfaces exposed to vacuum on either side of the slab. Density functional theory (DFT) is the preeminent tool for computing the energies (including surface energies) of inorganic solids. Typical DFT calculations can be used to optimize a surface structure and produce a corresponding energy at 0 K in vacuum. The resulting low-energy surface structures have shown good alignment with experimental measurements (e.g., using low-energy electron diffraction, LEED) of carefully prepared materials in near-vacuum conditions.[7–9]

For real systems and applications, the temperature and environment (e.g., gas composition) play significant roles in dictating the structure and energetics of inorganic surfaces. This motivates the application of different thermodynamic potentials for computing the energies. The DFT-calculated total internal energy ($E$) is a reasonable approximation for the enthalpy of a solid (because the pressure-volume contribution is small).[10] Mapping these enthalpies to Gibbs energies with first-principles calculations is much more computationally intensive because this requires computing the vibrational (phonon) free energies for all involved solids (including the slabs).[11] A common approximation in computational surface science is that the entropic contribution of the involved gaseous species (e.g., $O_2$ in air) is much larger than the entropic contribution from the involved solids.[12] Thus, a typical approach is to compute grand canonical surface energies for slabs allowed to exchange species with their environment as:

$$\gamma = \frac{1}{2A} * (E_{slab} - NE_{bulk} - \sum_i \Delta N_i \mu_i) \qquad [2]$$

where **Equation 1** is amended to account for the excess ($\Delta N > 0$) or deficiency ($\Delta N < 0$) of some species, $i$, in the slab compared with the bulk at chemical potential, $\mu_i$. In **Figure 2**, we illustrate that when species $i$ is gaseous (e.g., $O_2$), the effect of both temperature, $T$, and gas concentration, $p_i$, is captured using $\mu_i = \mu_i(T, p_i)$.[12]

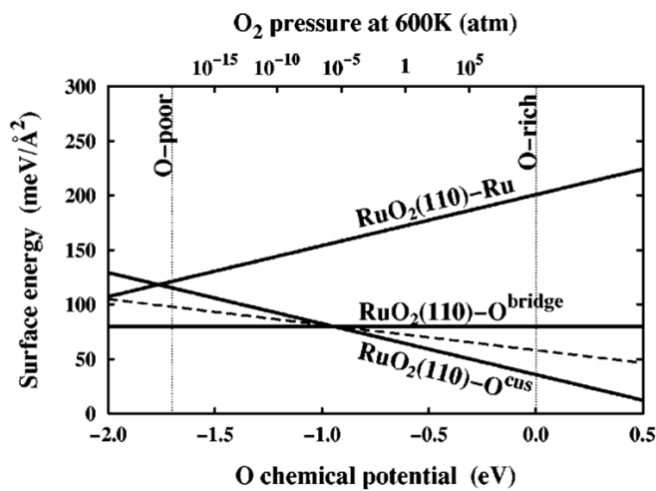

**Figure 2.** Surface free energies, $\gamma(T, p_{O2})$, for three possible $RuO_2(110)$ terminations calculated over the allowed range of oxygen chemical potential, $\mu_O(T, p_{O2})$, as indicated by the vertical dashed lines. The sloped dashed line depicts the surface free energy of a $RuO_2(110)$-$O^{cus}$ termination with only every second $O^{cus}$ site occupied. Reproduced from Ref. 12.

This approach can also be generalized to more complex thermodynamic environments (e.g., aqueous electrochemical environments using the Pourbaix potential).[13,14] An important consideration for the purposes of computational surface science is that these open systems introduce additional complexities as the surface composition (termination) can vary substantially depending on the temperature and environment.

While the aforementioned challenges are true for any particular facet (various terminations, restructuring), a further complication is that it is often critical to know the relative energies of many possible facets. Consider the Wulff construction, a prevalent method used to determine the equilibrium shape of a crystal of fixed volume, which is calculated by minimizing the total Gibbs free energy of the proposed system.[15,16] The minimization is performed with respect to the weighted product of facet surface energies and facet surface areas. As such, changes in the relative energies of the facets (i.e. due to changes in temperature or environment) can manifest as modifications to the equilibrium crystal shape. In **Figure 3**, we show how the computed equilibrium morphology of $RuO_2$ nanoparticles changes due to the dependence of relative surface energies on the change in oxygen chemical potential, $\Delta\mu_O$.[17]

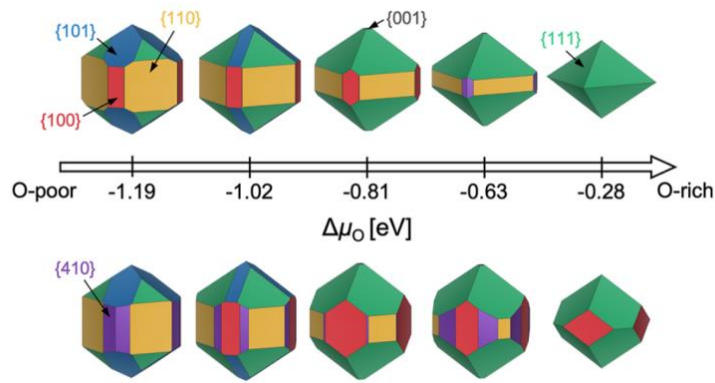

**Figure 3.** Equilibrium Wulff nanoparticle shapes computed from $RuO_2$ surface energies of all low-index (up to (111)) facets and the (410) vicinal, using locally optimized structures **(top)** and global geometry optimized structures **(bottom)**. The indicated oxygen chemical potentials, $\Delta\mu_O$, correspond to calcination pretreatment conditions used by Rosenthal et. al.,[18,19] Jirkovsky et. al.,[20] Lee et. al.,[21] and Narkhede et. al.,[22] from left to right. Standard conditions (300 K, 1 bar, -0.28 eV) are also displayed. Reproduced from Ref.[17] with permission from the American Chemical Society.

In an effort to cull the number of required calculations, many efforts have focused on a single facet[12,23–28] or a (sub)set of low-Miller index facets (e.g., up to (111)).[29–31] For some systems, this has led to good agreement with experimental measurements. For example, Reuter and Scheffler investigated the stability of the O-terminated, Ru-terminated, and stochiometric $RuO_2$(110) facets as a function of $\Delta\mu_O$, ranging from -2.0 eV to 0.5 eV.[12] They computed that a transition from the $RuO_2$(110)-$O^{cus}$ termination to the $RuO_2$(110)-$O^{bridge}$ termination, where *cus* and *bridge* refer to specific locations of oxygen on the surface, occurs at $T = 450 \pm 50$ K and $p_{O2} = 10^{-12\pm2}$ atm. This agrees with temperature desorption spectroscopy (TDS) measurements that found an excess of $O^{cus}$ atoms on the $RuO_2$(0001) surface at temperatures between 300-550 K under ultra-high vacuum ($p < 10^{-12}$ atm) conditions.[32] $RuO_2$(0001) has been found to form $RuO_2$(110) domains under oxidizing conditions.[7,33] The study of the $RuO_2$ system was extended by Wang et. al. to include all possible (1×1) terminations of the (100), (001), (110), (101) and (111) facets over the same range of $\Delta\mu_O$.[30] These surface energies were used to compute equilibrium particle morphologies as a function of $\Delta\mu_O$. The particle morphologies were qualitatively compared to scanning electron microscopy (SEM) images of experimentally grown $RuO_2$ nanoparticles,[18,19,22] where the major features (overall shape, facet coverage) of the computed morphologies were found to agree with experiment. For other systems, the inclusion of only low-Miller index facets can be a substantial approximation, and many facets that are relevant to the application of a material can be missed by only looking at this subset. In the case of Pd and Rh, Mittendorfer et. al. computed equilibrium particle morphologies with the inclusion of the (100), (110), (111), (211), (311), and (331) facets.[34] They discovered that under UHV conditions a significant fraction of the nanoparticle surface is comprised of the high-Miller index surfaces (211), (311), and (331).

So far, we have discussed that surface structures of interest can be generated as inputs to DFT calculations, which perform a local relaxation of the structure and yield accurate estimates for the internal energies. Using the thermodynamic relations discussed previously, these internal energies can be mapped to more useful thermodynamic potentials. However, an intrinsic limitation of this approach is that the only surface structures that can appear in the resulting surface phase diagrams are those that were specified as inputs by the user. Enumerating all possible surfaces (facets, terminations, reconstructions) and computing their energies with DFT is intractable. This motivates the development of sampling approaches to rationally explore the landscape of plausible surface structures. These approaches make use of concepts from crystal structure prediction,[35,36] optimization,[37] statistical mechanics,[38,39] and molecular dynamics simulations[40] (among other techniques). A detailed description of these methods is outside the scope of this review, but the application of machine learning (ML) in the context of these methods will be discussed. The remainder of this review will focus on the role of ML methods in facilitating accurate predictions of inorganic surface structures and energies under thermochemically relevant conditions.

## Machine learning interatomic potentials

The computational cost of energy evaluations with DFT scales approximately with the cube of the number of electrons in the system. This scaling means DFT calculations are often restricted to small numbers of structures, structures with less atoms, and very short timescales for molecular dynamics (MD). Interatomic potentials (IPs) are often used as surrogates for DFT and can scale approximately linearly with the number of atoms in the system. Historically, empirical IPs assume a particular functional form and fit parameters using higher fidelity data (e.g., from DFT) for some structures of interest. ML has recently emerged as a powerful tool for learning the relationship between crystal structures and DFT-calculated energies (and forces) that result. So-called machine learning interatomic potentials (MLIPs) have achieved remarkable performance as surrogates for DFT.[41–43] For bulk crystals, there have been demonstrations of "universal" MLIPs that are trained to perform well on materials spanning the periodic table.[44–47] Similarly, the Open Catalyst Project,[48,49] a massive open data challenge, has shown that MLIPs trained on millions of structures relevant to heterogeneous catalysis can be effective surrogates for predicting the structures and energies of surfaces with adsorbates.[50–56] Predicting the thermochemical stability of solid-state surfaces presents a different challenge, and MLIPs have not yet been shown to be effective "universal" surrogate models for solid-state surfaces. There have, however, been several examples of MLIPs dramatically accelerating the determination of surface phase diagrams within targeted materials spaces of interest.

A thorough review of MLIPs is outside the scope of this work, so we will briefly introduce two classes of MLIPs that have been applied extensively for surface science. The first approach relies upon Gaussian Process Regression (GPR) to develop so-called Gaussian Approximation Potentials (GAPs).[57–60] A typical procedure for fitting a GAP is shown in **Figure 4**. Briefly, the method begins by collecting ground-truth energies and forces (usually from DFT) for structures of interest to populate a database of reference data. For efficient training, these crystal structures must be "represented" in a manner that maximizes the retention of information subject to common invariances and equivariances that should be exhibited by periodic crystals.[61] GPR is then used to fit a probabilistic relationship between the target properties (energies, forces) in the reference data and the descriptors that result from the chosen representation.[58] Once the model is trained, any new configuration (structure) of interest can be represented in the same manner and passed through the model to infer the energies and forces associated with that structure. GAPs can also be used as the "force field" to drive MD simulations. Because the underlying model (GPR) is probabilistic, the resulting uncertainties can be used to iteratively improve the model using active learning.[62]

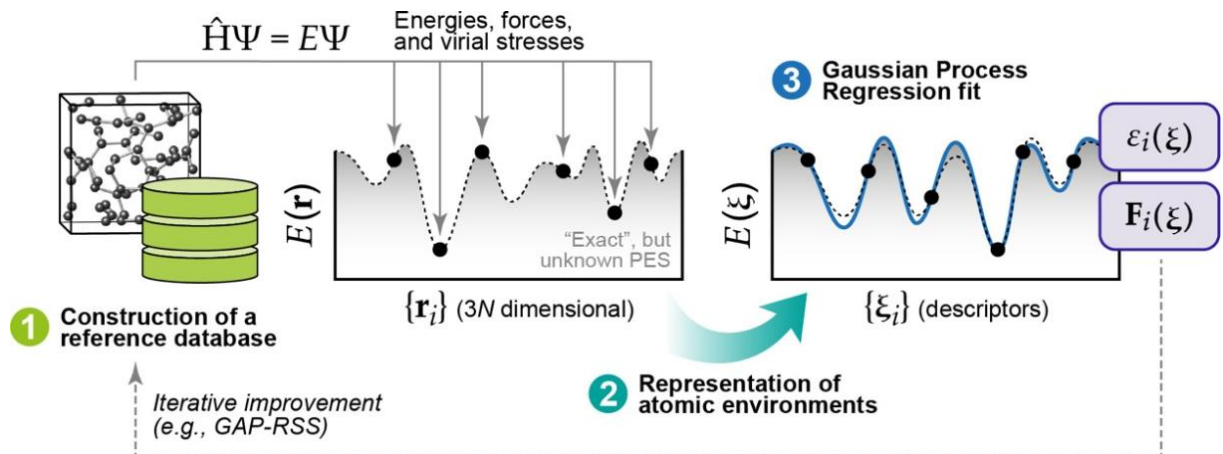

**Figure 4.** Three main components required for GAPs: **(1)** a robust reference database of quantum-mechanical data (usually generated with DFT), **(2)** a representation of the atomic environments associated with each reference point, and **(3)** the GPR model fit. Reproduced from Ref.[58] with permission from the American Chemical Society.

Alternative MLIP fitting approaches and architectures make use of many of the same concepts (reference data, representing crystal structures, model training, active learning), but may vary the underlying model and associated structural representation. As one example, the GPR model can be replaced with a deep learning model in the form of neural network (NN) potentials. As one class of NN potentials, graph neural networks (GNNs) leverage a graph representation for each crystal structure, where each node is an atom in the structure and neighboring atoms (within some radial cutoff distance) are connected via edges.[44–46,52] Aside from graphs, other well-known neural network potentials represent the crystal structure through equivariant descriptors, such as radial functions that are applied to and summed over distances from a central atom.[63–67] There are many flavors of NN potentials and the interested reader is encouraged to see more thorough reviews of MLIPs.[41,68–73] In the following sections, we review recent efforts to use MLIPs at various steps in the computational surface science pipeline.

## Direct predictions of surface energy

MLIPs are capable of rapidly and directly predicting the surface energy of a given slab, provided they have been appropriately trained for the material system of interest. This approach enables the accelerated exploration of a selected materials system with the potential to more comprehensively understand the energetics that may be missed using only DFT. With the goal of more robustly exploring possible $IrO_2$ surface structures, Timmermann and Reuter trained a GAP using 136 DFT-calculated structures.[40] The training data included 78 low-index facets, 34 bulk structures, and 20 nonequilibrium surface structures taken from high temperature MD simulations of various nanoparticle sizes and shapes. The GAP predicts that reordered (101) and (111) (1×1) structures are most stable under simulated annealing conditions (ramping to $T$ = 1000 K over 20 ps followed by slow cooling at 3 K/ps for 250 ps). This was further confirmed by DFT calculations as well as LEED and scanning-tunneling microscopy (STM) of annealed $IrO_2$ crystals. These results show how data-driven approaches can be leveraged to identify important surface structures that may have been missed using typical low-throughput approaches.

After previously identifying missed stable $IrO_2$ structures, Timmermann et. al. employed active learning in a two-stage framework for training GAPs to predict low-index surface structures of $IrO_2$ and $RuO_2$.[74] An initial GAP model was trained on DFT-calculated energies of $O_2$ dimers with varying O-O bond lengths, bulk unit cells of $MO_2$ ($M$ = Ir, Ru) at varying compressed, expanded, and optimized lattice parameters, and 21 low-Miller index (1×1) surfaces with $M$-, O-, stochiometric-, or peroxo-terminations. Sixteen of the low-Miller index surfaces, excluding the peroxo-terminations, were used as starting configurations for simulated annealing to generate 80 additional stable $IrO_2$ structures and 63 $RuO_2$ structures. The generated candidates were relaxed using DFT to assess differences in the GAP-predicted structures, which were measured as a function of the minimal similarity between two atoms within a structure, given by the Smooth Overlap of Atomic Positions (SOAP) kernel.[67] For those GAP-predicted structures where there were significant differences, the DFT-relaxed structure was computed and used for training in place of the GAP-predicted structure. The authors ultimately identified 8 $IrO_2$ and 7 $RuO_2$ terminations that are more stable than terminations formed by cleaving the bulk oxides for -2.0 eV < $\Delta\mu_O$ < 0 eV. In **Figure 5**, we show 8 of these novel terminations compared to their conventional bulk cleaved counterparts.

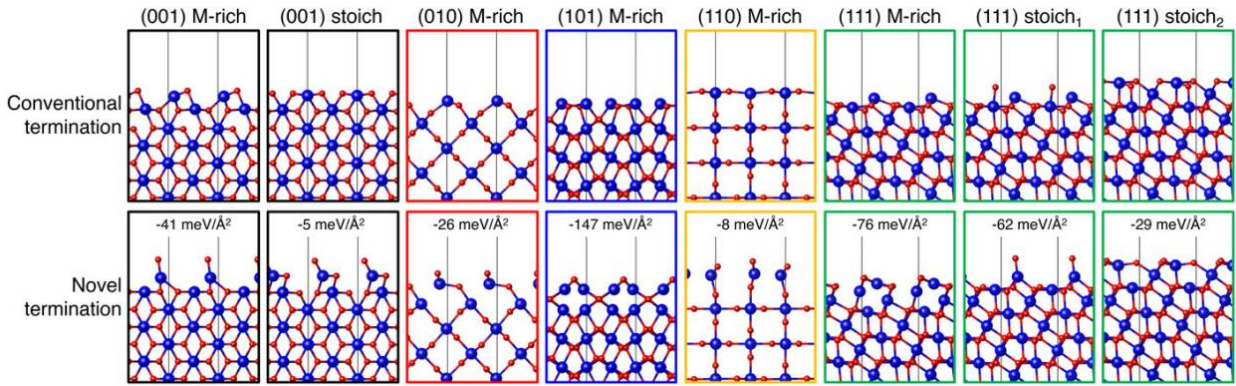

**Figure 5.** $IrO_2$ (1×1) surface structures identified with DFT (conventional) or during GAP training and surface exploration (novel). The top row depicts a side view of the conventional terminations resulting from bulk truncation and DFT geometry optimization. The bottom row depicts a side view of the GAP identified most stable structure, with the relative difference in surface free energy stated explicitly. Ir atoms are drawn as larger blue spheres and O atoms are drawn as smaller red spheres. Reproduced from Ref.[74] with permission from the American Institute of Physics.

Similar objectives have also been pursued using NN-based MLIPs rather than GAPs. Phuthi et. al. used data from 4548 structures (bulk, bulk with defects, pristine surfaces, and surfaces with adsorbates) generated through the DPGen active learning framework[75] to train NequIP[76] and Deep Potential[65] models for elemental Li.[77] Surface energy and nanoparticle morphology predictions were compared directly to DFT calculations, as well as predictions from a popular modified embedded atom (MEAM) empirical potential[78] and spectral neighbor analysis potential (SNAP)[64]. The authors show that both their NequIP and Deep Potential models achieve accuracies within 1 meV/Å$^2$ of the surface energy computed by DFT for higher-Miller index facets (up to (332)) despite their models only explicitly using the (100), (110), and (111) facets as starting structures for the active learning framework.

Similarly, Gao and Kitchin constructed a NN potential for Pd using the Atomistic Machine-learning package (Amp).[79,80] The NN architecture consisted of 2 hidden layers with 18 nodes each and was trained on ~2,700 DFT-calculated energies of bulk, slab, and defect structures. For the fcc(111) surface, the average surface energy was computed for supercells of size (2×2), (2×3), (3×3), (3×4), and (4×4). The average surface energy predicted by the model was in close agreement with DFT-computed average surface energies, with a mean absolute error (MAE) of < 2 meV/Å$^2$. Additionally, the surface vacancy energy was computed with DFT and the NN, where the authors found the NN to underestimate the DFT value by as much as 222 meV/atom, suggesting further tuning for defective surfaces would be needed. It is worth noting that the authors also compared the single point run time between DFT and their NN and found that the NN scaled linearly with the number of atoms and, on average, was four orders of magnitude faster than DFT.

## From surface energies to nanoparticle morphologies

We have so far discussed the speed and accuracy with which GAPs and NN potentials are capable of directly evaluating surface energies. If the relative surface energies among various facets and terminations can be predicted accurately, this enables the efficient prediction of equilibrium nanoparticle morphologies. Lee et. al. revisited the RuO$_2$ system to explore feasible surface reconstructions and compare DFT-calculated Wulff constructions with those of an updated GAP model.[17] Their updated model is an extension of the one previously trained by Timmermann et. al.[74] for the RuO$_2$ (1×1) surface structures and now includes RuO$_2$ c(2×2). The training for the new GAP potential added surface compositions with 25% and 75% additional oxygen coverage to the list of training data used for the initial (1×1) surface model. The inclusion of only 18 new surfaces with these new compositions enabled the model to predict critical reconstructions involving tetrahedral Ru$_{4f}$ motifs. The authors further utilized the GAP model to predict surface energies over the range $-1.5$ eV $< \Delta\mu_O < 0$ eV and computed the resulting equilibrium nanoparticle shapes. They noted that their particle morphologies resulting from GAP-predicted surface energies are qualitatively consistent with those reported by Wang et al., who calculated equilibrium shapes from surface energies computed strictly using DFT.[30] However, Lee's computed morphologies, shown in **Figure 3**, display a non-trivial percentage of the equilibrium particle morphology that is covered by the high-Miller index (410) facet, which was not shown by the previous low-Miller index studies.

Returning to NN potentials, Shrestha et. al. computed particle-size dependent phase diagrams as well as equilibrium particle morphologies for molybdenum and tungsten carbides.[81] Similar to Gao and Kitchin,[80] the authors utilized Amp[79] to develop separate NN potentials for each carbide system. The training was performed using DFT-computed energies for a total of 5918 Mo-C and 5941 W-C structures. The 5918 Mo-C structures included 154 Mo metal, 49 bulk (Mo$_x$C$_y$), and 5715 slabs with facets up to (111) and 49 high-Miller index facets for which the authors could find literature references. The 5941 W-C structures included 167 W metal, 46 bulk (W$_x$C$_y$), and 5728 slabs with facets up to (111) and 38 high-Miller index facets for which the authors could find literature references. Using these models, the authors predicted the surface energies of 1509 Mo$_x$C$_y$ and 1080 W$_x$C$_y$ surfaces up to Miller index 5 before generating Wulff constructions for $-0.5$ eV $< \Delta\mu_C < 0$ eV. For facets found in the equilibrium nanoparticles at various points in the $\Delta\mu_C$ range, the surface energy was computed using DFT. These DFT-computed surface energies of the NN-identified facets were then used to re-compute equilibrium particle morphologies. The resulting nanoparticle morphologies compared qualitatively well to transmission electron microscopy (TEM) and X-ray diffraction (XRD) measurements and are shown in **Figure 6** for the molybdenum carbide nanoparticles.[82]

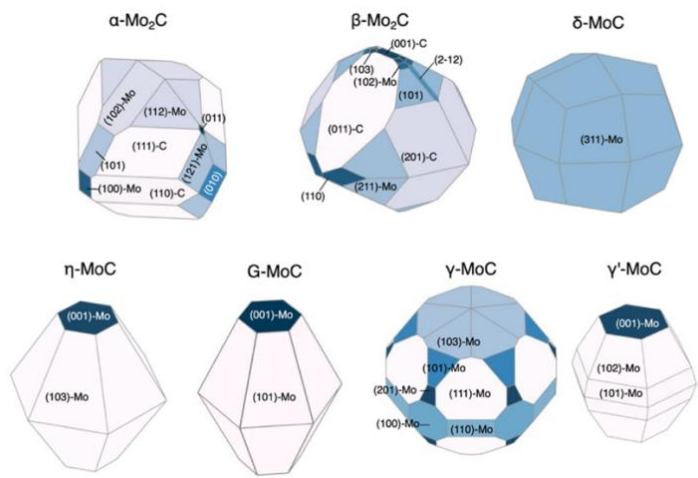

**Figure 6.** DFT-computed Wulff constructions of the equilibrium particle morphologies of different molybdenum carbide phases at $\Delta\mu_C = -0.15$ eV using NN-identified facets. Reproduced from Ref.[81] with permission from the American Chemical Society.

Leveraging direct predictions of surface energies is not the only method of predicting equilibrium nanoparticle morphologies. Palizhati et. al. utilized a crystal graph convolution neural network (CGCNN) to predict cleavage energies, or the energy required to break bonds along a specific plane, of bimetallic surfaces from which they compute Wulff constructions.[83] The cleavage energies are equal to the surface energies provided that the terminations of the resulting slabs are identical. The CGCNN was trained on cleavage energies of 3033 intermetallic surfaces spanning 36 different elements. The training cleavage energies were computed using a linear extrapolation method, where the total DFT-computed slab energy was plotted as a function of slab thickness, and the cleavage energy is given by the $y$-intercept. The authors assessed their model's accuracy by comparing the DFT-calculated Wulff constructions with the CGCNN-predicted Wulff constructions for NiGa, CuAl, and CuAu. They show that their model predictions of equilibrium particle morphology capture the majority of high area facets, with a highest area fraction MAE in the case of CuAu (MAE = 0.096).

## Energy inputs to Monte Carlo simulations

Surface reconstruction can lead to complex equilibrium geometries under changing temperatures or environmental conditions, which drastically affect final surface properties. When a single facet is of particular interest, more extensive sampling of the feasible surface structures can lead to an improved understanding of the relative surface energies. Such extensive sampling leads to more realistic predictions of the final observed structure but comes with the drawback of significantly higher computational cost and is typically intractable when very many facets are relevant (e.g., in Wulff constructions).

Sampling strategies are often based on Monte Carlo methods that can be used to explore the plausible reconstructions of a given surface under varying conditions. The rapid exploration of feasible reconstruction events is dependent on the speed and accuracy of the underlying surface energy calculator. Recently, Du et. al. developed a high-throughput active learning framework, Automatic Surface Reconstruction (AutoSurfRecon), for end-to-end prediction of surface energetics and exploration of surface reconstructions.[84] Their framework introduced a Virtual Surface Site Relaxation-Monte Carlo (VSSR-MC) method in the canonical and semi-grand canonical ensembles, which the authors showed can reproduce well known surface reconstructions of GaN(0001) (see **Figure 7a**) and Si(111). Following the demonstration of VSSR-MC, the authors mapped a phase diagram for $SrTiO_3$(100), shown in **Figure 7b**. For the calculation of the $SrTiO_3$(100) surface energies, the authors trained a neural network force field using the PaiNN[56] architecture. The predicted surface energies over the range -10 eV < $\Delta\mu_{Sr}$ < 0 eV yielded a double layer $TiO_2$ termination at low (more negative) $\Delta\mu_{Sr}$ and single layer $TiO_2$ to single layer SrO terminations at increasing $\Delta\mu_{Sr}$, all of which have been experimentally reported.[85–91] The authors computed the phase diagram of $SrTiO_3$(100) by also predicting the surface energies over the range -10 eV < $\Delta\mu_O$ < 0 eV. The authors note that their predicted phase diagram is qualitatively similar to that which was computed through DFT by Heifets et. al.[92]

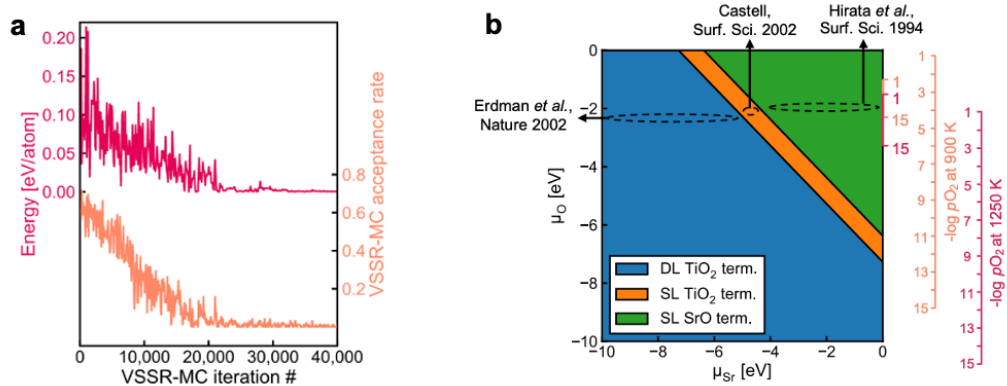

**Figure 7. (a)** A typical VSSR-MC run profile is depicted for high-temperature annealing of GaN(0001). **(b)** The NN-computed phase diagram of $SrTiO_3$(100) showing the stable surface terminations at varying $\mu_{Sr}$ and $\mu_O$ along with estimated positions of three experimental $SrTiO_3$(001) surfaces, Erdman et. al.,[87] Castell,[86] and Hirata et. al.[85] Four vertical axes are illustrated on the right. The smaller axes provide an abbreviated view of the larger axes. Reproduced and adapted from Ref.[84]

The previous investigation of surface reconstructions chose to avoid the computationally more expensive grand canonical Monte Carlo (GCMC), though in situations such as the study of oxidation processes, it may be necessary to use GCMC as it does not limit the interactions between the surface lattice and adsorbates. Therefore, Xu et. al developed a general framework for training NN potentials to be used with GCMC for exploring surface oxidation.[93] They tested the framework by exploring the $PtO_x$ system. 52,448 DFT-computed energies were used to train an Embedded Atom Neural Network Potential (EANNP)[94,95] to predict the surface and oxygen adsorption energies of the (111), (211), and (322) facets. Monte Carlo simulations were carried out using the EANNP and resulted in the discovery of formation mechanisms for the raised $PtO_4$, minimal stripe $Pt_2O_6$, and edge $PtO_6$ units, which were verified through replication by DFT calculations.

Boes and Kitchin took a slightly different approach for predicting oxygen absorption on Pd surfaces. They utilize the Amp package[79] to train a Behler-Parrinello (BP) NN[96] for the Pd(111) surface.[97] Their training data consisted of DFT calculations for 107 unique energy configurations of a 3x3x4 Pd slab. For each configuration, oxygen was placed at either the fcc, hcp, bridge, or top sites prior to relaxation. The authors used each step of the DFT-relaxation trajectories to provide 11,925 training data points to the model. GCMC was then performed with the BPNN as an energy calculator, where the authors predicted the relative potential energy barriers associated with oxygen migration across the Pd slab surface. Their results found good agreement with DFT and experimental energies, within 0.15 eV at any given site or nearest neighbor distance.[98–100] Boes and Kitchin note that the BPNN could be expanded for use in ternary systems of interest, leading to Yang et. al. training individual BPNNs for Pd, Au, and Cu.[101] Training was performed on 5100 DFT-computed surface energies of fcc(111) slabs with random compositions of the three elements. The individual BPNNs were then combined to predict surface properties of the ternary Cu-Pd-Au fcc(111) alloy. MC simulations were performed across 24 bulk compositions to explore metal segregation at the fcc(111) surface. The framework was able to qualitatively depict trends in the AuPd, and CuAu portions of the ternary space though it falls short in predicting the CuPd portions when compared to cluster expansion results.[102] The authors attribute this limitation to the use of ideal fcc(111) surfaces in generating their training data, as when fcc(110) surface data was incorporated the model was able to more consistently reproduce the CuPd behavior.

## Alternative ML-based sampling strategies

So far, we have discussed the implementation of MLIPs to enable accurate equilibrium particle morphology estimation and efficient probabilistic simulation. The training of the described MLIPs has largely focused on structures generated through domain knowledge, literature surveys, or automated active learning approaches. The following section is set to introduce recent works in sampling more robust training sets through less conventional search approaches. The focus is again on those that leverage ML, though other sampling strategies (e.g., nested sampling[39,103,104]) have also been used.

Zhu et. al. returned to the well-studied $RuO_2$ system to explore the structure of $Ru/RuO_2$ interfaces.[105] They used stochastic surface walking (SSW)[106] to generate more than $10^7$ (cluster, layered, and bulk) Ru-C-H-O structures. SSW is a Metropolis Monte Carlo[107] based search method that smoothly manipulates a given structure to generate new configurations. DFT-computed internal energies for 46,731 select structures were used for training a NN potential. The authors used a modified version of the phenomenological theory of martensitic crystallography[108] to generate plausible $Ru/RuO_2$ interfaces before optimizing the atomic coordinates and predicting the interfacial energies with their NN. The five most stable interfaces are shown in **Figure 8**. Three of the five most stable interfaces were matched with previous experimental results: $RuO_2(101)$ on Ru(1010), $RuO_2(101)$ on Ru(0001), and $RuO_2(100)$ on Ru(1010).[8,109] The SSW-NN framework facilitated Chen et. al. to develop an automated search for optimal surface phases (ASOP) in the grand canonical ensemble.[110] The SSW-NN method was used to generate 50,131 (cluster, layered, and bulk) Ag-C-H-O structures and train a NN for exploring the surface oxide phases of Ag(111) and Ag(100). The authors reproduced the experimentally observed Ag(111) c(4×8),[111] Ag(111) p(4×4),[112–114] and Ag(100) $(2\sqrt{2}\times\sqrt{2})R45°$[115–117] surface structures together with unreported, but predicted-low-energy, Ag(111) (2×1) and Ag(100) $(2\sqrt{2}\times2\sqrt{2})R45°$ surfaces.

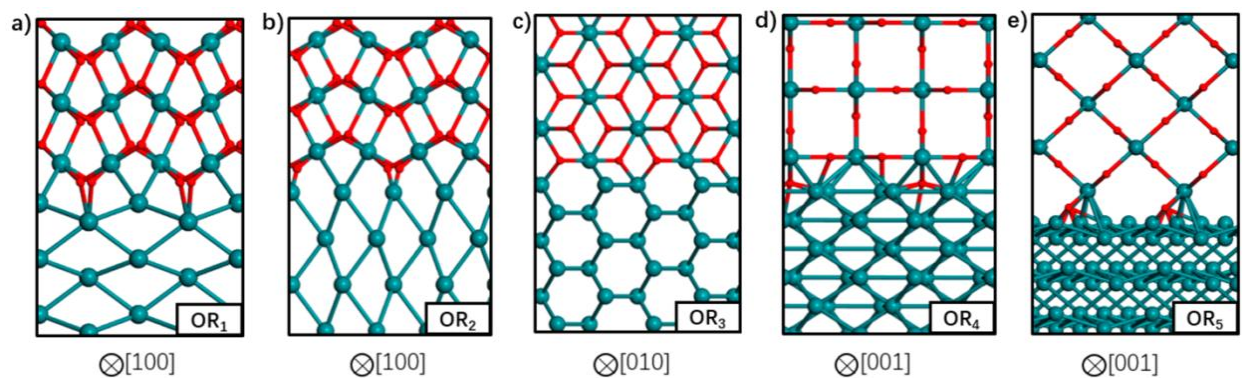

**Figure 8.** Atomic structures of the five most stable Ru−RuO$_2$ interfaces with different orientation relationships (OR). Ru atoms are depicted by the green balls and O atoms by the red balls. The crystallographic direction in RuO$_2$ bulk is indicated below each interface. Reproduced from Ref.[105] with permission from the American Chemical Society.

Evolutionary strategies have also been utilized to generate candidate training structures. To explore possible TiO$_x$ overlay structures on SrTiO$_3$, Wanzenbock et. al. combined the covariance matrix adaptation evolution strategy (CMA-ES),[118] which iteratively generates new overlay structures by perturbing existing surface atoms based on a normal distribution, and a NN potential.[119] The NN was trained on 3000 DFT-computed surface energies for overlayer structures generated by CMA-ES with SrTiO$_3$(110) (4×1) as the starting structure. The authors then performed a set of 50 CMA-ES runs using each of SrTiO$_3$(110) (3×1), (4×1), and (5×1) as initial structures and the trained NN potential as the energy calculator. This approach generated stable SrTiO$_3$(110) (3×1) overlay structures where TiO$_4$ tetrahedron create six- and eight-membered rings, shown in **Figure 9**, that were found to be consistent with STM images.[120] Additionally, the SrTiO$_3$(110) (4×1) seeded runs predicted six- and ten-membered rings of corner-sharing TiO$_4$ tetrahedron, also observed by STM images.[120–122] Finally, the SrTiO$_3$(110) (5×1) seeded runs predicted an STM-observed six- and twelve-membered ring structure[120] and a previously unobserved higher-energy eight- and ten-membered ring structure.

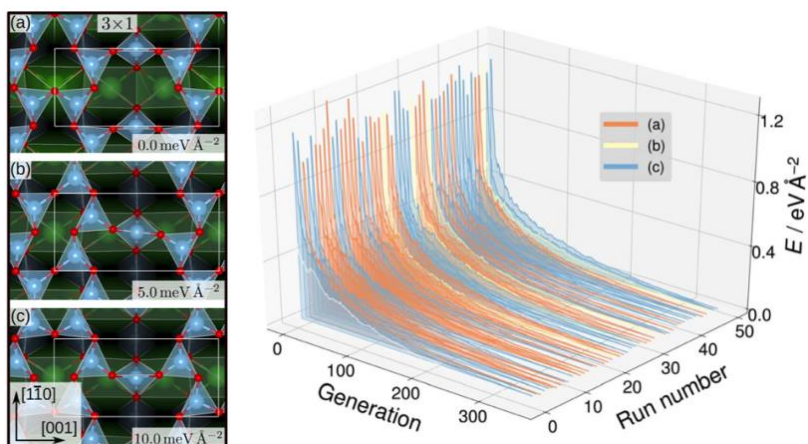

**Figure 9.** SrTiO$_3$(110) (3×1) reconstruction overlayers **(left)** identified by performing sets of NN-backed CMA-ES runs and further refined by two subsequent optimizations. Structures show corner-sharing TiO$_4$ tetrahedra in six- or eight-membered rings. The calculated energy minimum **(a)** is set to zero and the relative energies of the other arrangements, **(b)** and **(c)**, are shown. The energy trajectories of the 50 CMA-ES runs **(right)** on the SrTiO$_3$(110)

(3×1) surface, with the calculated energy minimum set to zero. The labels **(a)**, **(b)**, **(c)** correspond to the overlayers shown on the left. Reproduced and adapted from Ref.[119] with permission from the Royal Society of Chemistry.

In pursuit of further advancing evolutionary search approaches, Bisbo and Hammer developed the global optimization with first-principles energy expressions (GOFEE) strategy, which generates new candidate structures by perturbing atoms in a subset of structures from an initial population.[123] The new candidate structures are relaxed using a GPR model, initially trained on a user-selected set of relevant structures. An acquisition function is used to assess which of the generated structures to select for DFT single-point energy evaluation. After evaluation, the structure is added to the initial training dataset and the process is repeated. In this way, the GPR model is improved while simultaneously exploring the energy landscape. The authors tested their method by reproducing a well-known $SnO_2(110)$ (4×1) surface reconstruction, observed both experimentally[124] with LEED and computationally[125] using evolutionary algorithms. Bisbo and Hammer also explored the intercalation of oxygen between graphene grown on Ir(111), an experimentally well-studied process.[126–128] The authors find that an oxidized graphene edge lifts slightly from the Ir(111) surface, which may allow for intercalation. In pursuit of further improving the efficiency of GOFEE, Merte et. al. modified the strategy to update the training set with subsets of the generated structures instead of a single structure.[129] This improved strategy was used to explore the surface structure of $Pt_3Sn(111)$ with a (4×4) oxygen overlay. In conjunction with STM, LEED, X-ray photoelectron spectroscopy (XPS) and low-energy ion scattering (LEIS) data,[130,131] the authors were able to propose and validate a surface composition of $Sn_{11}O_{12}$ and surface structure with Sn in 3-fold coordination with oxygen.

With the expansion of efficient search algorithms, a compact and flexible framework for training set generation and model production could further accelerate the development of accurate MLIPs. Here, Christiansen et. al. developed the atomistic global optimization X (AGOX) package,[132] which allows users to build their own dataset-generation pipelines based on flexible modules for performing random-structure search, basin-hopping, evolutionary-structure generation, and GOFEE.[123,129] The AGOX package was built to train GPR models as energy calculators. The versatility of the package has been demonstrated by Ronne et. al. who trained an Ag GPR model based on the SOAP[67] representation by implementing parallel basin-hopping, which generates new structures using a stochastic perturbation of atoms in an initial structure.[133] Twelve concurrent basin-hopping searches were performed from starting overlay structures with compositions $Ag_xO_y$ ($x$=4, 5, or 6 and $y$=2, 3, 4, or 5) on Ag(111). The concurrent searches generated structures that were subsequently fed into a shared database and used to train a single GPR. The model reproduced the stable Ag(111) c(4×8) structure.[110]

Aside from evolutionary and stochastic searches, alternative attempts applied learning strategies from fields such as computer vision for improving surface structure searches. Jorgensen et. al. developed an atomistic structure learning algorithm (ASLA) that leverages convolutional neural networks (CNN) and reinforcement learning to construct 2D and planar structures atom-by-atom.[134] Within reinforcement learning, a model is required to make decisions based on an expected "reward", such as maximizing a chosen function. The ASLA is split between three stages: building, evaluation, and training. The building stage involves the placement of atoms by the model one-by-one on a real space grid to generate a structural candidate. The placement is restricted by a minimum distance between atoms and dictated by the CNN, which predicts the expected "reward" received by each atom placement. Within the ASLA framework, the reward is the minimization of the internal energy of a candidate structure, where the true energy is computed by DFT during the evaluation stage. The CNN is then updated based on the root mean square error between the expected energy of the generated structure and the DFT-computed energy. Through this iterative approach, the model learns to "build" structures of minimal energy without prior knowledge of the system of interest, at the potential cost of preforming many DFT calculations. The underlying grid that the structure is built upon can be empty or populated by atoms (e.g., for building overlay structures on a specific facet). The authors demonstrated the capabilities of this approach by building the p(4×4) oxygen overlay structure on an underlying Ag(111) surface, which was reproduced from experimental observation by the ASOP framework as discussed previously in this review.[110] Meldgaard et. al. expanded the ASLA framework to 3D predictions of surface reconstructions by increasing the dimensionality of the CNN. The method was verified by reproducing the minimum energy anatase $TiO_2$(001) (1×4) reconstruction, as observed by STM imaging.[135] Meldgaard then demonstrated the ability to apply transfer learning within the ASLA approach by reproducing the LEED-observed and DFT-predicted $SnO_2$(110) (4×1) reconstruction,[124] starting from the generation of stable $SnO_2$(110) (1×1) reconstructions.

## Conclusions and Perspective

Throughout the previous sections, we have reviewed the application of MLIPs for modeling inorganic surfaces. In several places, high-throughput or automated structure generation, model training, and analysis workflows were pivotal (e.g., DP-GEN,[75] Amp,[79] AutoSurfRecon,[84] ASOP,[110] AGOX[132]). Automated and publicly available frameworks have been a key aspect of accelerating the understanding of equilibrium particle morphologies and surface reconstruction mechanisms under different environments. Systematic workflow development has continued to be a focus of the community with examples including a recent semi-autonomous workflow, WhereWulff,[136] which takes as input a stable bulk structure and performs the necessary bulk truncation, first-principles calculations, and surface optimization to compute Wulff constructions, generate Pourbaix diagrams, and preform reactivity analysis. Other examples exist for producing physics-based potentials[137] performing model finetuning,[138,139] and further exploring surface reconstructions[140] bringing improved functionality to the fingertips of those working on surface science. In addition to lowering the barrier of entry for newcomers to this field, these (semi-)autonomous frameworks also enable the magnitude of systematic data generation required for efficient model training. These large, systematic datasets make open data repositories paramount for managing and compiling the generated data in a common format to foster more rapid model training and development and avoid duplication of efforts. Several projects including OCP,[48,49] Crystalium,[141–143] Colabfit,[144] and NOMAD[145] have begun to fill such roles for subsections of the surface science community. Even with open access to the data required to train MLIPs, exhaustive sampling (particularly in large-scale systems) becomes intractable due to computational costs.[43] This motivates a push to further accelerate energy evaluations (e.g., lower the inference time of MLIPs).[146] Beyond directly predicting the phase stabilities of inorganic surfaces, MLIPs open up new possibilities to explore complex problems such as materials synthesis prediction (where nanoscale effects may be important)[13,147–149] and catalyst degradation (which may involve a complex traversal of many surfaces).[150] Overall, the continued improvement of MLIPs with more data, better model architectures, improved sampling strategies, and reduced inference times promises to open new possibilities for the computational modeling of inorganic surfaces.

## Acknowledgments

The authors gratefully acknowledge support from the University of Minnesota in the form of new faculty start-up and a student fellowship from the College of Science & Engineering Data Science Initiative.